\documentclass[reprint,amsmath,amssymb,aps,]{revtex4-2}
\usepackage{etoolbox}
\usepackage{subfig}
\usepackage{graphicx}
\usepackage{dcolumn}
\usepackage{bm}
\usepackage{hyperref}
\usepackage[braket, qm]{qcircuit}
\usepackage{algorithm}
\usepackage{algorithmic}

\begin{document}

\preprint{APS/123-QED}

\title{Quantum-Enhanced Simulation-Based Optimization \\for Newsvendor Problems}

\author{Monit Sharma$^{1}$}
\author{Hoong Chuin Lau$^{1,2}$}
\email{Corresponding author email: hclau@smu.edu.sg}
\address{$^1$School of Computing and Information Systems,  Singapore Management University, Singapore}
\address{$^2$Institute of High Performance Computing, A*STAR, Singapore}

\author{Rudy Raymond$^{3,4}$}
\address{$^3$IBM Quantum, IBM Japan, Japan}
\address{$^4$Dept.~of~Computer~Science,~The University~of~Tokyo,~Japan}

\begin{abstract}

Simulation-based optimization is a widely used method to solve stochastic optimization problems. This method aims to identify an optimal solution by maximizing the expected value of the objective function. However, due to its computational complexity, the function cannot be accurately evaluated directly, hence it is estimated through simulation. Exploiting the enhanced efficiency of Quantum Amplitude Estimation (QAE) compared to classical Monte Carlo simulation, it frequently outpaces classical simulation-based optimization, resulting in notable performance enhancements in various scenarios. In this work, we make use of a quantum-enhanced algorithm for simulation-based optimization and apply it to solve a variant of the classical Newsvendor Problem which is known to be NP-hard. Such problems provide the building block for supply chain management, particularly in inventory management and procurement optimization under risks and uncertainty. 
\end{abstract}

\maketitle

\section{\label{sec:level1}Introduction}
Mathematical optimization methods are vital for tackling various real-world business management problems. They have been widely used in fields such as logistics, supply chain management, and finance to find the best solutions that reduce costs, improve efficiency, and achieve other important goals. Traditional optimization methods have proven to be effective in solving a wide range of problems. Some common types of mathematical optimization problems include linear programming (LP), integer programming (IP), mixed-integer programming (MIP), quadratic programming (QP), and non-convex programming, among others. These problems are applied in various areas, from optimizing supply chains to managing portfolios and fine-tuning machine learning models.

Many optimization methods operate under the assumption of complete knowledge about problem data or its underlying probability distribution. However, real-world situations often present data that is subject to uncertainty. Stochastic optimization addresses this inherent uncertainty by integrating probabilistic elements into the optimization model formulation. One approach to solving stochastic optimization problems is Simulation-Based Optimization or SBO (see for example, \cite{SBO}), which has garnered significance due to its effectiveness in handling complex systems influenced by uncontrollable factors such as weather conditions or significant events like conflicts or pandemics. Among various SBO methods, we focus on the Sample Average Approximation (SAA) method, which is a popular technique that approximates the expected value of an objective function involving uncertain parameters via a sample average of realizations of the uncertain parameters.

Monte Carlo simulation represents a similar approach to solving stochastic optimization problems (e.g. \cite{monte}). Both Monte Carlo simulation and SBO rely on generating random samples to approximate solutions and assess outcomes. Monte Carlo simulation focuses on analyzing system behavior under varying input conditions, while SBO is concerned with setting decision variables to achieve specific objectives, such as maximizing profit or minimizing costs. Unlike Monte Carlo simulation, SBO typically seeks optimal solutions by adjusting decision variables iteratively across multiple simulation runs.

It is worth highlighting that both SBO and Monte Carlo simulation can become computationally demanding, particularly when dealing with high-dimensional optimization problems. Their computational load is often characterized by the number of samples denoted as $M$ (unrelated to the system's dimension). The estimation error in Monte Carlo simulation scales as $\mathcal{O}(M^{-1/2})$ \cite{monte}. This implies that the error, which represents the disparity between the estimated outcome and the true value, diminishes proportionally to the square root of the number of samples. In other words, doubling the number of samples reduces the error by approximately a factor of $1/\sqrt{2}$, i.e., one would need four times as many samples to halve the error. The convergence rate is governed by the law of large numbers, with the caveat that this convergence slows down as the number of samples $M$ grows.

Quantum computing taps into the principles of quantum mechanics to execute intricate information processing tasks \cite{qcqi}, ushering in a fresh era of problem-solving methods. This shift in approach has spawned inventive strategies for tackling a diverse range of challenges, spanning domains such as quantum chemistry \cite{qchem}, optimization \cite{qopt}, machine learning \cite{qml}, and finance \cite{portfolio,quantmonte}. Quantum Amplitude Estimation (QAE) \cite{QAE} is a quantum algorithm that achieves $\mathcal{O}(M^{-1})$ convergence, providing a quadratic speedup over classical Monte Carlo \cite{QAE}, and can be applied across various problem domains, akin to Monte Carlo simulations. Recent endeavors have delved into its utility in tasks like risk analysis \cite{risk} and option pricing \cite{option}.

Apart from QAE, other quantum optimization techniques like the Variational Quantum Eigensolver (VQE) \cite{vqe} and the Quantum Approximate Optimization Algorithm (QAOA) \cite{qaoa} are mainly suited for addressing Quadratic Unconstrained Binary Optimization (QUBO) problems. These methods have their limitations when applied to optimization problems in different categories that are not easily translatable into an Ising Hamiltonian representation.

In contrast, QAE offers a more extensive application scope, encompassing a wider range of optimization challenges. QAE shows promise in significantly speeding up SBO for both continuous and discrete variables. This quantum-driven enhancement introduces a novel approach referred to as Quantum-Enhanced Simulation-Based Optimization (QSBO) \cite{qsbo} which is suitable for handling objective functions characterized as expectation values, variances, cumulative distribution functions (CDFs), or (conditional) value at risk ((C)VaR) of random variables or functions derived from them. It is worth noting that all these metrics can be viewed as expectation values.
This approach has also been used to tackle a concise yet highly practical issue in the field of supply chain management, with a specific focus on the Newsvendor Problem. These implementations are constructed using Qiskit \cite{qiskit} framework.

In this paper, our contributions are as follows:

We utilize the maximum profit formulation for the Newsvendor Problem, which has broader applicability compared to the minimal loss formulation commonly found in the literature, though mathematically equivalent, one formulation can be easily converted into the other. This formulation allows us to address the NP-hard non-linear extension of the problem \cite{or}. Secondly, previous research in this field commonly relied on a known demand distribution, often represented by a normal distribution with a specified mean and standard deviation ($\mathcal{N}(\mu,\sigma)$). In contrast, our approach involves an unknown demand distribution, where a function may not precisely capture the demand in such settings, to address this, we employ Quantum Generative Adversarial Networks (qGANs) \cite{qgan} to load the unknown demand distribution, thereby creating a more realistic scenario.
Thirdly, in terms of quantum optimization technique, we improve the simulation-based optimization method in \cite{qsbo} by introducing a new comparison operator, thereby reducing the number of qubits needed in the circuit (see Appendix \ref{comparisonop}).
\\
Lastly, we validate the effectiveness of our approach experimentally by conducting experiments using both the $\textit{ibmq qasm simulator}$ and the noisy fake backend of $\textit{ibmq singapore}$ and the training of qGAN was done on the actual quantum backend of $\textit{ibm torino}$. We implement error mitigation techniques and compare the results to assess the performance of our method.

The subsequent sections of this paper are structured as follows: Section \ref{sec:level2} provides a comprehensive discussion of the Newsvendor Problem, and what makes the non-linear extension of it NP-hard to solve. Section \ref{sec:level3} explains the details of Quantum Amplitude Estimation and what makes it a viable method to solve stochastic optimization problems that require the calculation of an expectation value. In Section \ref{sec:level4}, we delve into the Quantum-Enhanced Simulation-Based Optimization (QSBO) algorithm and discuss how it can solve for both discrete and continuous variables. Section \ref{sec:level5} outlines the construction of the required quantum circuit for the problem and presents the implementation of the Newsvendor Problem within this framework, with error mitigation techniques. Section \ref{sec:level7} presents the experimental results of our model. Finally, in Section \ref{sec:level8}, we provide concluding remarks and explore potential avenues for future extensions and research.

\section{\label{sec:level2} Newsvendor Problem (NV)}

A fundamental single-period problem in stochastic optimization is the Newsvendor Problem (NV). A vendor needs to decide how many units ($x$) of an item with a short life cycle (such as newspapers) to order based on the demand distributions, the costs of ordering, and the revenue from the sales. 

Let the cost of ordering $x$ units at the beginning of the period be $c(x)$, and the revenue of selling $x$ units throughout the period be $r(x)$, where all these functions are non-negative and non-decreasing. The stochastic demand $D$ for the item is a discrete random variable contained in $[0,..., M]$. 

Arrow et al. \cite{Darrow} showed that if all costs and revenues are linear i.e $c(x) = cx, r(x) = rx$ (these cost parameters must satisfy $r>c$, otherwise the problem can be trivially solved), they have a simple solution (see Appendix \ref{form}).

Most studies of the NV assume that the vendor is provided analytical representations for the revenue, cost functions, and demand distributions. This assumption does not always align with real-world scenarios. 
In some cases, suppliers refrain from revealing the order cost function $c(\cdot)$ to the vendor and instead provide cost quotations $c(x)$ in response to each query $x$ from the vendor. Such situations are frequent in spot markets, as seen in scenarios like a tourism agency booking a block of seats on a specific flight. It is unrealistic to anticipate the airline to disclose the detailed function $c(\cdot)$ that dictates seat allocation across various booking classes. The particular focus is on scenarios where demand forecasting is crucial.

For demand distributions, organizations often maintain databases containing historical customer demand data, which is periodically updated with point-of-sale data, typically on a daily or weekly basis. In such contexts, the demand distribution does not follow a statistical distribution (such as normal distribution), instead, the database provides probabilistic information regarding the likelihood of demand (or sales) falling below a specified threshold for any given value examined by the user.

In \cite{or}, the authors are concerned with problems where the demand distribution is given by an oracle rather than a fixed distribution. It established that certain variants of the NV require exponential queries to the oracle and proved that a certain class of NV problems are NP-hard. This holds even when the revenue functions are linear, and the cost functions contain fixed as well as linear components.

While solving the linear NV to find the optimal order quantity is relatively straightforward, the computation of the expected profit is exponential in the worst case. This underscores a pronounced disparity in the computational complexity between determining the decision variable (i.e., the order quantity) for the optimal solution and evaluating its corresponding value (i.e., profit). Importantly, this result's proof also implies that addressing the NV with cost functions composed of fixed and linear elements necessitates an exponential number of queries to find both the optimal order quantity and the associated expected profit.

The non-linear NV poses a formidable challenge as it lacks a general constant factor approximation algorithm. The complexity arises from the fundamental observation that finding the minimum of an arbitrary integer-valued function $f:[0,....., N] \rightarrow \mathbb{Z}$, or determining whether it occurs in the intervals $[1,.....,\lfloor N/2 \rfloor ] $ or $[\lfloor N/2 \rfloor,....N ] $, requires $N+1$ queries in the worst-case scenario. This underscores the intricate nature of the problem and highlights the significant computational effort needed to arrive at optimal solutions.

Although, if we have additional information about the function, the number of queries might be reduced. For instance, if the function is monotone, only two queries are required. If the function is convex, only $\mathcal{O}(\log N)$ queries are needed. But if the function is unimodal with a unique minimum, the number of queries needed is $N$. See \cite{or} for detailed proof.

In this paper, we are interested in the following Newsvendor Problem that requires an exponential number of function evaluations, which are all NP-hard whenever all cost functions and the CDF/PDF are all computable in polynomial time:

\begin{itemize}
    \item Calculating the order quantity that maximizes expected profit, even if the revenue function is linear and ordering cost is fixed plus linear.
\end{itemize}

In classical methods such as classical Simulation-Based Optimization (SBO), the runtime complexity can grow exponentially with the problem size in the worst case. However, Monte Carlo methods are frequently used in classical solving, we illustrate how Quantum Simulation-Based Optimization (QSBO) can improve the performance of the Monte Carlo algorithm by achieving the desired result with fewer samples.

More precisely, we focus on the maximum profit formulation for the NV problem, where we have the objective function $f(s,d)$, the revenue from it is linear and the cost of procurement is fixed plus linear:

\begin{equation}\label{eq:2}
    f(s,d) = \text{min}(s,d) r - (s c + \delta(s\ge 1) t)
\end{equation}

where $r$ is the revenue cost per item, $c$ is the unit cost of purchasing per item, $t$ is the fixed price for placing an order, $d$ represents the demand and $s$ is the supply quantity (which are all discrete).

Effectively the problem is now to maximize the expectation value of the said function over all possible demand realization ${d_i} \in D$ :

\begin{equation}\label{eq:3}
    \mathbb{E}_{\text{D}} [\text{profit}] = \mathbb{E}_{\text{D}}\text{max}[f(s,d)]
\end{equation}

\section{\label{sec:level3}Quantum Amplitude Estimation (QAE)}

QAE, introduced in \cite{QAE}, is a quantum algorithm that provides a quadratic speedup compared to the traditional Monte Carlo simulations typically run on classical computers. It assumes that the problem of interest is given by a unitary operator $\mathcal{A}$ acting on $n+1$ qubits such that

\begin{equation}\label{eq:4}
    \mathcal{A}|0\rangle_n |0\rangle= \sqrt{1-a}||\Psi_0\rangle_n |0\rangle + \sqrt{a}|\Psi_1\rangle_n |1\rangle
\end{equation}

where $a \in [0,1]$ and $|\Psi_0\rangle$ and $|\Psi_1\rangle$ are two orthonormal states.

QAE enables the estimation of the parameter $a$ with a high probability, resulting in an estimation error that scales as $\mathcal{O}(1/M)$, where $M$ represents the number of $\mathcal{A}$ operator applications. The key component in this process is the Grover operator $\mathcal{Q}$, which is constructed as follows:

\begin{equation*}
    \mathcal{Q} = \mathcal{A} S_0 \mathcal{A^{\dagger}} S_{\Psi_0}
\end{equation*}
Here, $S_{\Psi_0} = \mathbb{I} - 2 |\Psi_0 \rangle \langle \Psi_0| \otimes |0\rangle \langle 0|$ and $S_0 = \mathbb{I} - 2|0\rangle_{n+1} \langle 0|_{n+1}$, as elaborated in detail in \cite{QAE}. The applications of $\mathcal{Q}$ are commonly referred to as \textit{oracle queries}.

\begin{figure}[h!]
    \centering
    \includegraphics[width = \linewidth]{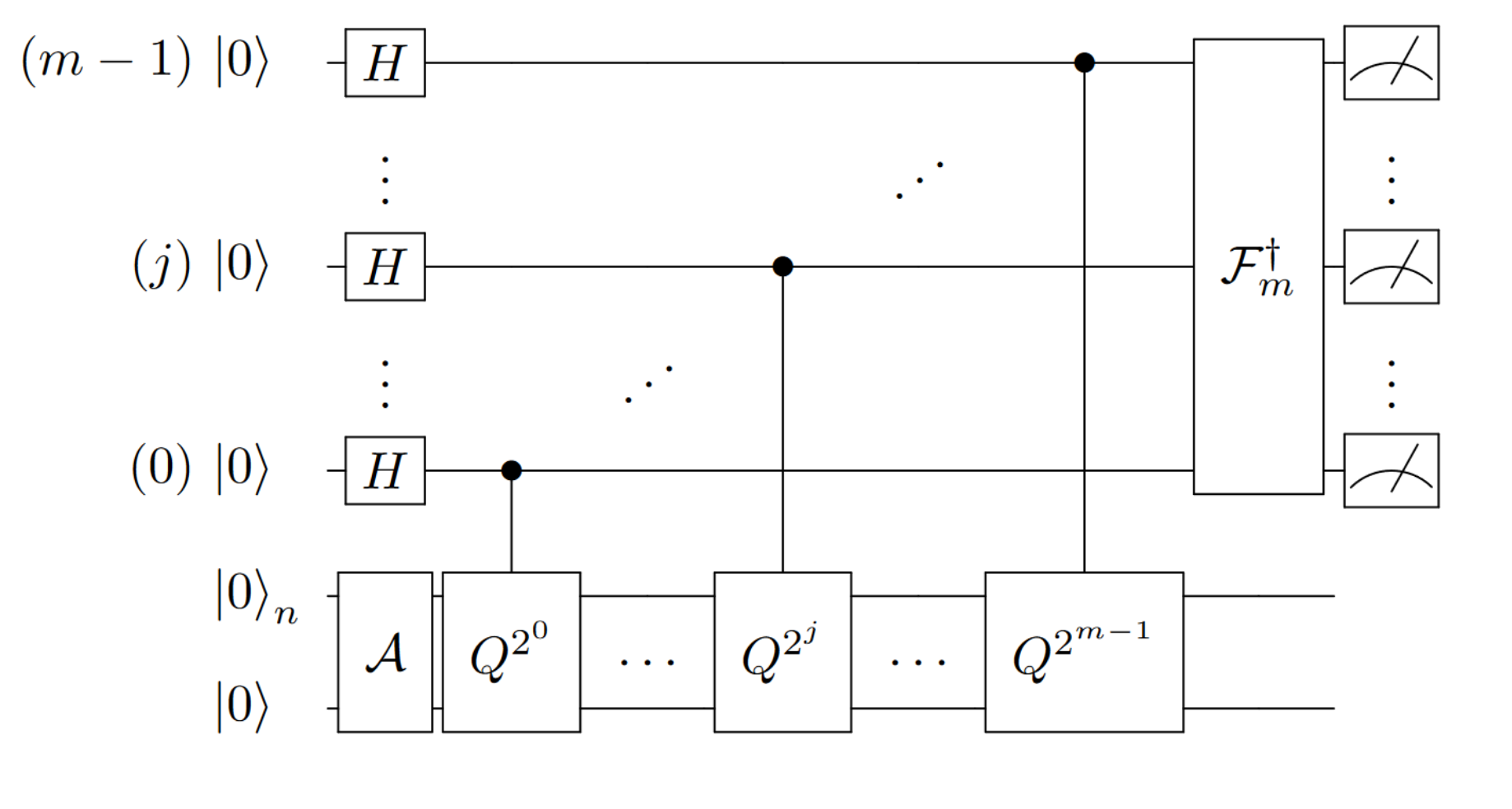}
    \caption{QAE circuit with $m$ ancilla qubits and $n+1$ state qubits. $H$ is the Hadamard gate and $\mathcal{F}^{\dagger}_m$ denotes the inverse Quantum Fourier Transform on $m$ qubits.}
    \label{fig:1}
\end{figure}

The standard form of QAE is derived from Quantum Phase Estimation (QPE) \cite{qpe}. It incorporates an additional set of $m$ ancillary qubits, all initially set in an equal superposition state. These ancillary qubits play a crucial role in representing the outcome. The number of quantum samples is defined as $M = 2^m$, and the algorithm involves the application of geometrically increasing powers of the $\mathcal{Q}$ operator, controlled by the ancillary qubits. Finally, it culminates in executing a Quantum Fourier Transform (QFT) on the ancillary qubits just before measurement, as depicted in Fig. (\ref{fig:1}).

Following this, the integer measurement result, denoted as $y$ and falling within the range ${0, \ldots, M-1}$, is converted into an angle $\tilde{\theta}_a$ using the formula $\tilde{\theta}_a = \frac{y \pi}{M}$. Subsequently, the estimated value of $a$, denoted as $\tilde{a}$, is defined as $\tilde{a} = \sin^2 (\tilde{\theta}_a)$, which lies within the interval $[0, 1]$. The estimator $\tilde{a}$ satisfies the following relationship:

\begin{align}\label{eq:5}
    |a -\tilde{a}| &\le \frac{2\sqrt{a(1-a) \pi}}{M} + \frac{\pi^2}{M^2} \\
    & \le \frac{\pi}{M} + \frac{\pi^2}{M^2} = \mathcal{O}(M^{-1})
\end{align}

with probability of at least $\frac{8}{\pi^2} \approx 81\%$. This illustrates a quadratic speedup compared to the $\mathcal{O}(M^{-1/2})$ convergence rate of classical Monte Carlo methods \cite{QAE}. By repeating this process multiple times and leveraging the median estimate, the probability of success can be substantially increased, approaching nearly $100\%$ \cite{risk}. These estimates $\tilde{a}$ are confined to the grid ${\sin^2 (y \pi/M) : y = 0,....,M/2}$ based on the potential measurement outcomes of $y$. Due to the symmetry of the sine function, the algorithm's output $\tilde{a}$ resides on a grid with $M/2 + 1$ possible values within the range of $[0,1]$.

The standard Quantum Amplitude Estimation (QAE) method often results in intricate quantum circuits and discrete estimates, denoted as $\tilde{a}$, contingent on the evaluation qubits $m$. Consequently, recent advancements have introduced several QAE variations to enhance both the accuracy and simplicity of the algorithm \cite{simple,raymond, counting, iter, faster}. These alternative methods provide a continuous range of estimated values while streamlining the circuit by eliminating the need for ancilla qubits and the Quantum Fourier Transform (QFT). Owing to their capability to generate estimates across a continuous spectrum, these QAE variations are better suited for Simulation-Based Optimization (SBO) tasks.

All variants of QAE, that do not make use of QPE are based on the fact that:

\begin{equation}\label{eq:7}
\begin{split}
    \mathcal{Q}^k \mathcal{A}|0\rangle_n |0\rangle =& \cos((2k+1)\theta_a |\Psi_0\rangle_n |0\rangle  + \\
    & \sin((2k+1)\theta_a)|\Psi_1\rangle_n |1\rangle
\end{split}
\end{equation}

where $\theta_a$ is defined as $a = \sin^2(\theta_a)$. In other words, the probability of measuring $|1\rangle$ in the last qubit is given by 

\begin{equation*}
    \mathbb{P}[|1\rangle] = \sin^2 ((2k+1) \theta_a)
\end{equation*}

The algorithms primarily differ in their strategies for determining the powers $k$ of the Grover operator ($\mathcal{Q}$) and how they consolidate the results into the final estimate of $a$. For our work, we will utilize the Iterative Quantum Amplitude Estimation (IQAE) algorithm \cite{iter}, and the Maximum Likelihood Quantum Amplitude Estimation (MLQAE) algorithm \cite{raymond}.

QAE is used to estimate an expected value $\mathbb{E}[f(X)]$, for a given random variable $X$ and a function $f: \mathbb{R} \rightarrow [0,1]$. Expected values in this form commonly appear as objective functions in Simulation-Based Optimizations (SBO).

\
For loading the probability distribution onto a quantum circuit, we make a quantum operator $\mathcal{P}_X$ that acts like 
\begin{equation}\label{eq:8}
    \mathcal{P}_X |0\rangle_n = |\psi\rangle_n = \sum_{i = 0}^{N-1} \sqrt{p_i} |i\rangle_n
\end{equation}
where the probability of measuring the state $|i\rangle_n$ is $p_i \in [0.1]$ with $\sum_{i=0}^{N-1} p_i = 1$ and $N = 2^n$. The state $|i\rangle_n$ is one of the $N$ possible realizations of a bounded discrete random variable $X$ which for instance represents a discretized demand for newspaper stocks. We load the discretized probabilities $p_i$ into the amplitude of $n$ qubits employing this operator $\mathcal{P}_X$.

Creating a qubit register to approximate a probability density function (PDF) can be accomplished through quantum arithmetic \cite{inter}, provided that the function is efficiently integrable. Alternatively, for all smooth and differentiable functions, this can be achieved using matrix product states \cite{smooth}. Another approach involves utilizing quantum generative adversarial networks (qGANs) to approximate generic functions \cite{qgan}. We will make use of qGANs to load our demand distribution.

We consider a function $f: \{0,...,N-1\} \rightarrow [0,1]$ and a corresponding operator:
\begin{equation}\label{eq:9}
    F|i\rangle_n |0\rangle = \sqrt{1-f(i)}|i\rangle_n |0\rangle + \sqrt{f(i)}|i\rangle_n |1\rangle
\end{equation}

 for all $i \in \{0,....,N-1\}$ acting on an ancilla qubit.

An operator like $F$ can be created through the utilization of quantum arithmetic and related methods \cite{efficient}, \cite{opquan}. In our research, we will adopt the approach outlined in \cite{risk}. This approach involves approximating $f$ using a Taylor expansion and utilizing controlled Pauli rotations to map the function values onto the qubit amplitudes. This mapping allows for the flexibility to strike a balance between precision and circuit complexity. This equilibrium can be attained by selecting an appropriate number of Taylor terms to approximate the function, thus avoiding the need for intricate quantum arithmetic operations.

For a discrete random variable $X$ taking values in $\Omega_X = \{x_i\}^{N-1}_{i=0}$, where $N= 2^n$ for a given $n$, with the corresponding probabilities $p_{x_i} = \mathbb{P}[X=x_i]$, the expectation value can be written as 

\begin{equation}\label{eq:10}
    \mathbb{E}[f(X)] = \sum_{i=0}^{N-1} p_{x_i} f(x_i) = \sum_{\hat{x} = 0}^{N-1} p_{\phi(\hat{x})} f(\phi(\hat{x})) 
\end{equation}

where $\phi : \{0,...,N-1\} \rightarrow \Omega_X$ represent the affine transformation from $\hat{x} \in \{0,...,N-1\}$ to $x \in \Omega_X$.

We can encode $\mathbb{E}[f(X)]$ in $\mathcal{A}$ making use of Eq. (\ref{eq:9}) and Eq. (\ref{eq:10}). First we load the discretized probabilities using Eq. (\ref{eq:8}) into the amplitudes of $n$ qubits by means of $\mathcal{P}_X$

Then we add one more qubit and use $F$ from Eq. (\ref{eq:9}) to define $\mathcal{A} = F(\mathcal{P}_X \otimes \mathbb{I})$. The state after applying it is given by:

\begin{equation}\label{eq:11}
\begin{split}
    \mathcal{A}|0\rangle_n |0\rangle  =&    \sum_{i=0}^{N-1} \sqrt{1-f(i)} \sqrt{p_i}|i\rangle_n|0\rangle \\
    +& \sum_{i=0}^{N-1} \sqrt{f(i)} \sqrt{p_i}|i\rangle_n|1\rangle 
\end{split}
\end{equation}

We use amplitude estimation to approximate the probability of measuring $|1\rangle$ on the last qubit, which implies that 

$$ a = \sum_{i=0}^{N-1} p_i f(i)$$
which is the desired expected value from Eq. (\ref{eq:10})

\section{\label{sec:level4}Quantum-Enhanced Simulation-Based Optimization}

Consider a discrete random variable $X$ with values $\in$ $\Omega_X$, a decision variable $y \in \mathbb{R}^d$, and a function $f$ mapping  $\Omega_X \times \mathbb{R}^d \rightarrow \mathbb{R}$. Our objective is to find:
$$ y^{*} \in \text{argmax}_y \mathbb{E}[f(X,y)]$$
To compute the solution, we use Quantum Amplitude Estimation (QAE) to calculate the expected value and apply classical optimization techniques to find an optimal solution, specifically denoted as $y^{*}$.

This decision variable $y$ may take on continuous forms, like setting the best product price or best location for a shop, or discrete forms, such as selecting the most suitable assets for a portfolio or determining the right stock article quantity to maintain.

For a discrete decision variable, we use $k$ qubits to represent $y$ with $2^k$ values. For instance, if $y \in \{0, 1\}^k$, we map $y$ to $|y_0, . . . , y_{k-1}\rangle_k$. The quantum state representing $y$ is parameterized, allowing an optimization routine to adjust the parameters and thereby search for an optimal solution $y^{*}$\cite{qsbo}. Drawing inspiration from VQE, this parameterization is implemented using a parameterized circuit $V (\theta), \theta \in \mathbb{R}^p$, to construct a trial state $|\psi(\theta)\rangle = V (\theta)|0\rangle.$

The selection of $V$ determines the Hilbert space that can be explored for the optimal solution $|y^{*}\rangle_k$, making it crucial for a successful implementation. Ideally, $V$ should cover a significant portion of the corresponding Hilbert space while maintaining a low circuit depth, meaning a minimal number of non-sequential circuit operations \cite{express}.

In VQE, this parameterized circuit serves as a trial wave function for estimating the ground state of a molecule. In the context presented here, it functions as a trial state for solving the problem at hand.

In other words, we define $|y(\theta)\rangle_k = V (\theta)|0\rangle_k$. The optimization problem can be formulated as finding the $\theta$ parameters that maximize the expectation value of our objective function,

$$ \theta^{*} \in \text{argmax}_{\theta} \mathbb{E}[f(X, y(\theta)]$$

such that for any given $\theta$ the result of QAE is based on a superposition $|y(\theta)\rangle$ over all possible solution candidates, i.e $y(\theta)$ is a random variable, instead of a fixed value.

The $\mathcal{A}$ operator consists of probability loading operator ($\mathcal{P}_X$) which loads the probability distribution of the random variable $X$ into the qubit register $|x\rangle_n$, the trial state $V (\theta)|0\rangle_k$ representing the decision variable $y$ in register $|y(\theta)\rangle_k$ and a function mapping $F$ acting on $|x\rangle_n$ as well as $|y(\theta)\rangle_k$.

$$ F|x\rangle_n |y\rangle_k |0\rangle = |x\rangle_n |y\rangle_k \big( \sqrt{1 - f(x,y)}|0\rangle + \sqrt{f(x,y)}|1\rangle\big)$$

This leads to 

$$ \mathcal{A}_{\theta}|0\rangle_{k+n+1} = F(V(\theta) \otimes P_X \otimes I) |0\rangle_k |0\rangle_n |0\rangle_1$$

A single assessment of the objective involves several steps: First, we prepare the operator $\mathcal{A}_{\theta} $, then utilize it to construct the operator $\mathcal{Q}_{\theta}$. Next, we execute the QAE procedure to obtain an estimation $\tilde{a}_{\theta}$ that approximates $\mathbb{E}[f(X,y)]$. By employing a classical optimization technique (like COBYLA), we can use the QAE output as an objective function to fine-tune the parameters for $y$ to obtain the optimal value $y^*$.

The conventional QAE method, as detailed in the initial part of Section \ref{sec:level3}, may pose difficulties for this type of optimization. According to the conventional QAE approach, an additional set of $m$ qubits is required to execute the Quantum Fourier Transform (QFT) operation, as part of the Quantum Phase Estimation (QPE) technique. When $m$ evaluation qubits are employed, the resulting estimate is confined to a fixed grid comprising $M/2+1$ points, resulting in a step-like function. Consequently, optimizing such a function can be challenging. Using other variations of QAE discussed earlier (like the IQAE, and MLQAE \cite{iter,raymond}) does not introduce such discontinuities and is a preferable choice.

\section{\label{sec:level5}Quantum Circuit}

Here we outline the building blocks needed to do the inventory management on a quantum computer. The critical components include the probability distribution describing the evolution of random variables $X$ on the quantum circuit, constructing the operator that computes the payoff function, and calculating the expectation value of the payoff. 

To compute the cost function of the Newsvendor Problem with QAE, we reformulate $f$ from Eq.  (\ref{eq:2}) as a conditional addition and subtraction.
The objective function changes to
\begin{equation}\label{eq:cost}
\begin{split}
    f(s,d)  = \delta(s\ge d) d r + & \delta(d \ge s) s r - \delta(s = d) s r - \\
    & s c - \delta(s\ge 1) t
\end{split}
\end{equation}

where $\delta(.)$ is a delta function and is $1$ if and only if the argument of the function holds and is $0$ otherwise, where $r$ is the revenue cost per item, $c$ is the cost of purchasing per item, $t$ is the fixed price while ordering, $d$ represent the demand and $s$ is the supply amount.

\subsection{Distribution Loading}
The initial component of our model is a circuit that operates on a probability distribution derived from historical sales data, treating it as the current demand. This circuit loads the distribution onto a quantum register, where each basis state signifies a potential value, and its amplitude corresponds to the associated probability. Given an $n-$qubit register, demand data $\{D_i\}$ for $i \in \{0,..,2^n -1\}$ and corresponding probabilities $\{p_i\}$, the distribution loading module creates the state:

\begin{equation*}
   |\psi\rangle_n = \sum_{i = 0}^{N-1} \sqrt{p_i} |i\rangle_n
\end{equation*}

The analytical formulas used in option pricing \cite{option} and Newsvendor \cite{qsbo} assume that the underlying data follows a log-normal distribution. In \cite{inter} the authors show that log-concave probability distributions can be efficiently loaded in a gate-based quantum computer. However, these simplified assumptions fail to capture important market dynamics, limiting the model's applicability in real-life scenarios. As such, the market-implied probability distribution of the underlying needs to be captured properly for valuation models to accurately estimate the intrinsic value.

The loading of arbitrary states into quantum systems requires exponentially many gates \cite{expo}, making it inefficient to model arbitrary distributions as quantum gates. Since the distributions of interest are often of a special form, the limitation may be overcome by using quantum Generative Adversarial Networks (qGANs) \cite{qgan}. These networks allow us to load a distribution using a polynomial number of gates. A qGAN can learn the random distribution $X$ underlying the observed data samples $\{x^0, x^1,...x^{k-1}\}$ and load it directly into a quantum state (see Appendix \ref{quantumgan}).

After the qGAN training, the generator gives us the parameters $\theta_p$ of a quantum state that represents the target distribution.

\begin{equation}\label{eq:12}
    |\psi(\theta_p)\rangle_n = \sum_{i=0}^{2^{n}-1} \sqrt{p_i }|i\rangle_n
\end{equation}

The $n-$qubit state $|i\rangle_n = |i_{n-1}....,i_{0} \rangle$ encodes the integer $i = 2^{n-1}i_{n-1} + ....+ 2i_1 + i_0 \in \{0,..,2^n -1\}$ with $i_k \in \{0,1\}$ and $k=0,....,n-1$. The probabilities $p_i$ approximate the random distribution underlying the training data. This approach can be easily extended to multivariate data, where we use a separate register of qubits for each dimension \cite{qgan}.

\subsection{Computing the Payoff}

We obtain the expectation value of a linear function $f$ of a random variable $X$ with QAE by creating the operator $\mathcal{A}$ such that $a = \mathbb{E}[f(X)]$ (see Eq. (\ref{eq:10})). Once $\mathcal{A}$ is implemented we can prepare the state in Eq. 
 (\ref{eq:4}), and the $\mathcal{Q}$ operator.

The payoff function for the Newsvendor Problem is piece-wise linear and as such we only need to consider linear functions $f: \{0,...,2^n-1\} \rightarrow[0,1]$, which we can write as $f(i) =f_1 i + f_0$. We can efficiently create an operator $F$ that performs Eq. (\ref{eq:9}), and making use of controlled $Y$-rotations give us:

\begin{equation}{\label{eq:13}}
    |i\rangle_n |0\rangle \rightarrow|i\rangle_n \big(\cos [f(i)]|0\rangle + \sin [f(i)]|1\rangle \big)
\end{equation}

To implement the linear term of $f(i)$ each qubit $j$ (where $j \in \{0,...,n-1\})$ in the $|i\rangle_n$ register acts as a control for a Y-rotation with angle $2^j f_1$ of the ancilla qubit. The constant term $f_0$ is implemented by a rotation of the ancilla qubit without any controls.

The operator $\mathcal{A}$ then, acts as 

\begin{equation}
\begin{split}
    \mathcal{A}|i\rangle_n |0\rangle = \sum_{i=0}^{2^n-1} \sqrt{p_i (\theta_p)} \cos [f(i)]|i\rangle_n |0\rangle \\ 
    +  \sum_{i=0}^{2^n-1} \sqrt{p_i (\theta_p)} \sin [f(i)]|i\rangle_n |1\rangle
\end{split}
\end{equation}

and is used within QAE. The output of QAE is then transformed to an estimate of the expectation value by reverting the applied scaling (see Appendix \ref{payoff}).

\subsection{\label{sec:level6}Circuit Optimization}
Quantum error mitigation techniques are widely employed to address and reduce errors that manifest during quantum computations on real hardware. Minimization of quantum circuit depth and width, along with the mitigation of gate errors, plays a pivotal role in enhancing the viability of circuit-based quantum computation for practical applications. Employing hardware optimization techniques is instrumental in diminishing errors, and improving precision in quantum computing outcomes.

Following a comprehensive analysis of various error sources, a selection of error mitigation techniques are used. Mapomatic \cite{mapomatic} is employed to optimize qubit mapping for specific quantum devices, Dynamic Decoupling \cite{dd} is utilized to mitigate qubit decoherence. These can be set using the optimization level and resilience level in qiskit transpiler.
\\

\section{\label{sec:level7}Results}
In this section, we illustrate the results of the QSBO method. Initially, we apply QSBO on a noise-free quantum simulator (\textit{ibmq qasm simulator}), subsequently, extending this model to a simulated noisy fake backend (\textit{ibmq singapore}), and employ error mitigation techniques to mitigate the noise. In our work, we make use of a bimodal distribution, the underlying function of which remains unknown. This distribution is encoded onto $n=3$ qubits and truncated within the range of $\Omega_D = [0,7]$. Considering that we can represent $2^n$ values using $n$ qubits, this framework can be seamlessly scaled to $\Omega_D = [0,2^n -1]$ by employing $n$ qubits to encode the demand value.

\begin{figure}[h!]
    \centering
    \includegraphics[width= \linewidth]{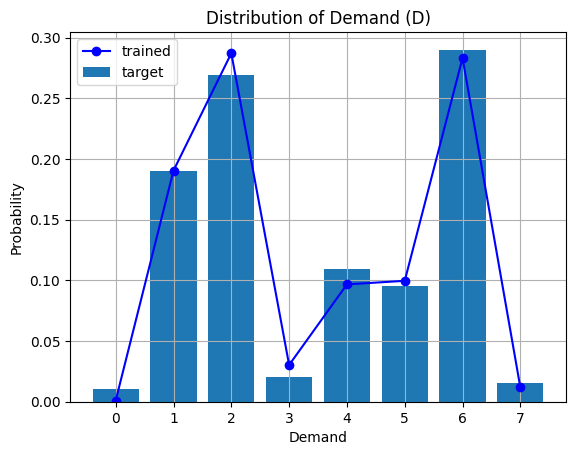}
    \caption{The demand distribution learned by the qGAN model(in line plot) as compared to the original demand distribution(in bar plot).}
    \label{fig:5}
\end{figure}

The distribution is fed into the quantum generator of the qGAN, where the model is trained to load a quantum circuit that transfers this distribution onto a quantum circuit. The optimization process employs COBYLA, with entropy loss acting as the loss function. The training process spanned $200$ epochs, with the optimizer learning rate of $10^{-3}$. Fig. (\ref{fig:5}) demonstrates the comparison between the qGAN-learned distribution and the original distribution.

\begin{figure}[h!]
    \centering
    \includegraphics[width = \linewidth]{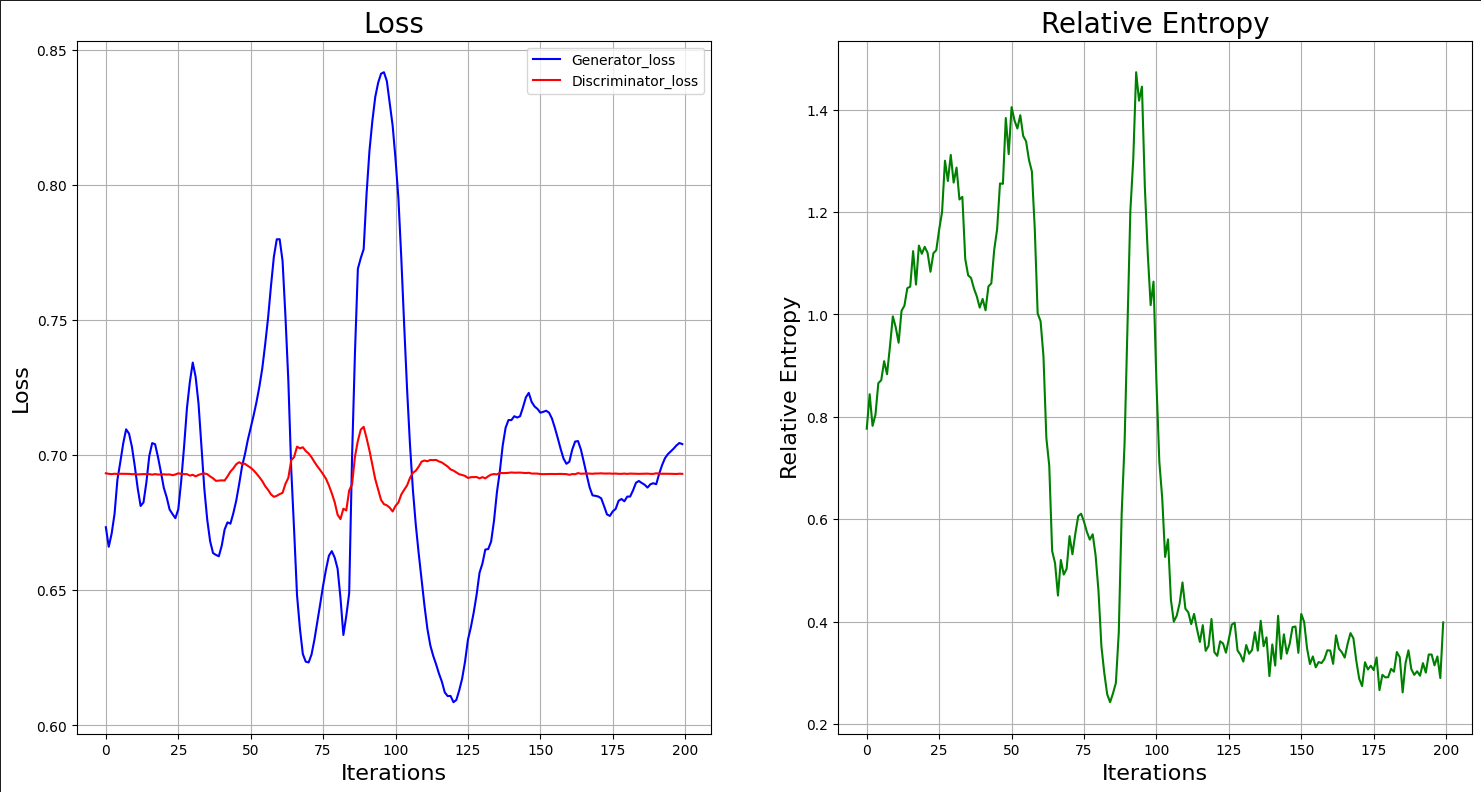}
    \caption{The progress in the loss and relative entropy during the training of the qGAN with \textit{ibm torino} quantum computer.}
    \label{entropy1}
\end{figure}

A comparison of the progress of the loss functions and the relative entropy for a training run is shown in Fig. (\ref{entropy1}). The relative entropy gradually converges to values close to zero, indicating that the learned distribution progressively aligns with the underlying training data samples. Additionally, the behavior of the loss functions provides further insights into the training dynamics and the optimization trajectory, offering valuable information for refining the training strategy and enhancing model performance.

For validating the use case, we will explore two scenarios: (i) maintaining a fixed demand distribution while varying the cost of procurement, and (ii) experimenting with different demand distributions while keeping the parameters for the cost of procurement constant.

\subsection{Fixed Demand and Varying Costs}

We made use of the trained qGAN demand shown in Fig. (\ref{fig:5}), and experimented on two different scenarios, (i) varying purchasing cost($c$) and (ii) varying fixed cost($t$).

\subsubsection{Varying Purchasing Cost}

Considering the cost function in Eq. (\ref{eq:cost}), we set the revenue cost per item as $r=0.6$ and the fixed cost as $t = 0.1$. We then adjust the cost of purchasing per item $c$, vary within the range of $[0.1,0.5]$.

\begin{figure}[h!]
    \centering
    \includegraphics[width= \linewidth]{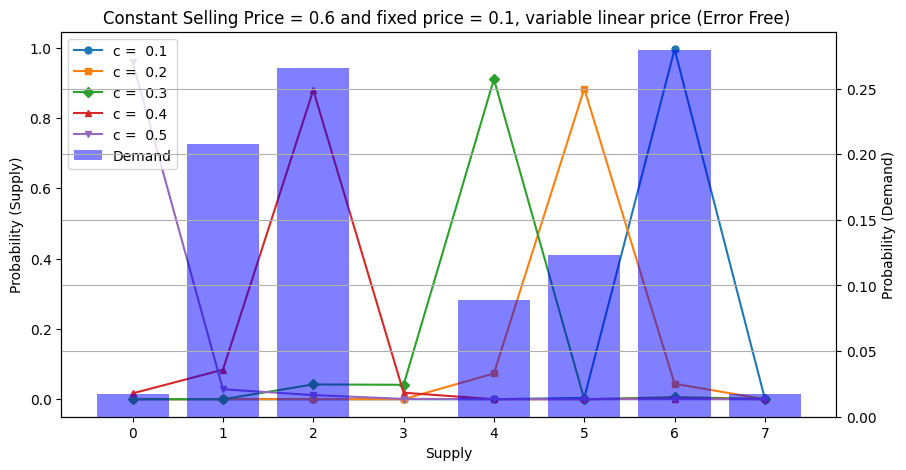}
    \caption{Optimal order quantity, for a fixed demand distribution (in bar plot) and various purchasing prices (in line plot) $c \in [0.1,0.5]$.}
    \label{fig:6}
\end{figure}

Fig. (\ref{fig:6}) illustrates a consistent demand distribution alongside fluctuating linear purchase prices.
The bar chart visualizes the demand distribution, with the corresponding probabilities shown on the right  $y$-axis. Additionally, the line graph depicts the optimal supply quantities, along with their probabilities on the left $y$-axis, for various purchase price values, and the $x$-axis represents the units of perishable goods (like newspaper batches in case of Newsvendor Problem).

 For instance, in the scenario where the cost of purchasing is $c= 0.2$, the supply peaks at $s=5$, suggesting an order of $5$ quantities. The result is shown in Fig. (\ref{fig:6}).

We see that as the purchasing cost escalates, the model tends to reduce its order quantity. This approach stems from the realization that while ordering more items may potentially boost sales, it could concurrently decrease profits. Consequently, as the purchasing cost converges toward the revenue cost, the model leans towards ordering a smaller quantity to maintain inventory levels. The primary objective remains profit maximization, rather than maximizing sales volume.

\subsubsection{Varying Fixed Cost}\label{subsec:vary}

\begin{figure}[h!]
    \centering
    \includegraphics[width = \linewidth]{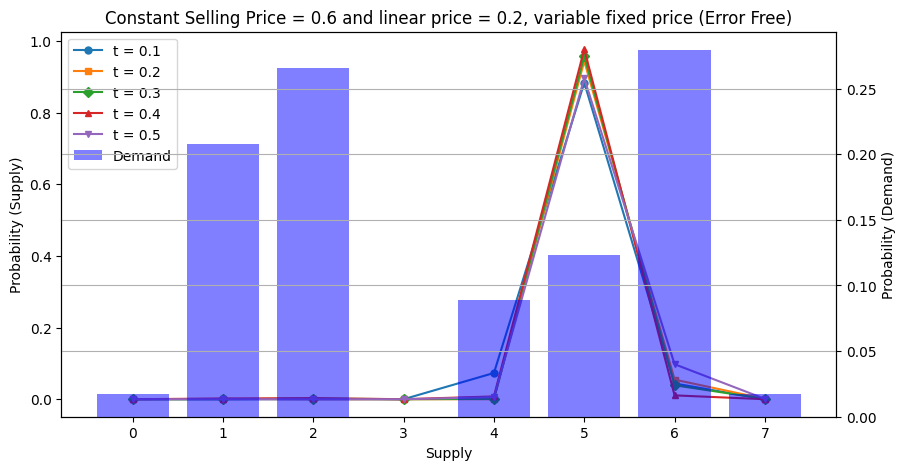}
    \caption{Optimal amount to order for varying fixed price of ordering $t \in [0.1,0.5].$}
    \label{fig:7}
\end{figure}

In this scenario, we assume a revenue cost per item of $r = 0.6$ and a cost of purchasing per item of $c = 0.2$, while varying the fixed cost, denoted by $t \in [0.1,0.6]$. The result is depicted in Fig. (\ref{fig:7}).

The optimal order quantity remains consistent regardless of the fixed price. While the fixed price does influence the profit value, the model consistently identifies the optimal order quantity that maximizes profit. Given that the purchasing price per item is constant, it remains a rational choice to order more items to maximize profit.

\subsection{Fixed Costs and Varying Demand }

In these experiments, we kept all the costs constant, we set revenue cost $r=0.6$, the cost of purchasing $c=0.3$, and the fixed cost of buying $t = 0.2$ and changed the respective demand distribution for two different scenarios. This approach allowed us to comprehensively assess the impact of varying demand patterns on our optimization model.

\begin{figure}[h!]
    \centering
    \includegraphics[width=\linewidth]{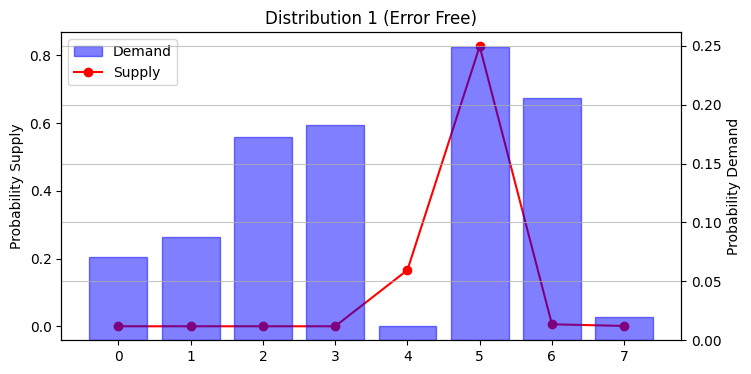}
    \caption{Optimal amount to order, for fixed cost parameters, and a different demand scenario. The $x-$axis represents the amount of goods, and the red line plot shows the probability of the amount to order.}
    \label{fig:8}
\end{figure}

In the first scenario, we observe a bimodal distribution, the model optimizes the ordering quantity, and the outcome is depicted in Fig. (\ref{fig:8}).

The resulting supply value is $5$, with over $80\%$ probability in the error-free simulator. This outcome aligns with our expectations from the model and corresponds closely to the classical result.

\begin{figure}[h!]
    \centering
    \includegraphics[width=\linewidth]{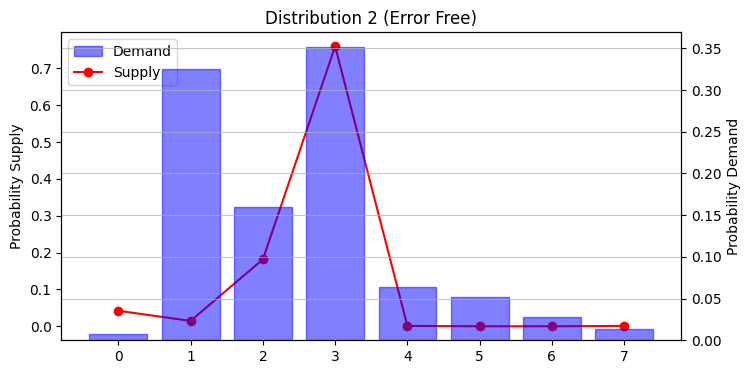}
    \caption{Optimal amount to order, for the second demand scenario, for a fixed set of cost. The $x-$axis represents the amount of goods with the red line plot representing the probability of the amount to order. }
    \label{fig:10}
\end{figure}

For the second distribution, featuring peaks at $1$ and $3$. Despite the relatively equal likelihood of both of these peak demand values, it remains intriguing to observe that the model consistently optimizes the one that leads to maximum profit. This pattern is evident in the results, as depicted in Fig. (\ref{fig:10}).

The model correctly identifies the maximum profit at $3$, indicating that ordering $3$ quantities will maximize our profit.

These findings show the robustness of the model in profit optimization across diverse demand scenarios. The consistency observed in quantum simulators highlights the reliability of the model in making profit-maximizing decisions, a crucial aspect of real-world business applications. Additionally, the alignment between quantum and classical results emphasizes the reliability of quantum simulation in accurately optimizing supply decisions amidst complex demand scenarios.

\subsection{Error Free vs Error Mitigated Results}

Since the experiment involving the QSBO algorithm was conducted on two distinct computational backends, it is worth noting the differences in their outcomes. The first backend, characterized by its error-free simulation environment, produced results that closely mirrored the expected outcomes from classical computations as reported in the results above. This alignment between quantum and classical results serves as a testament to the algorithm's effectiveness under idealized, noise-free conditions.

\begin{figure}[h!]
    \centering
    \includegraphics[width = \linewidth]{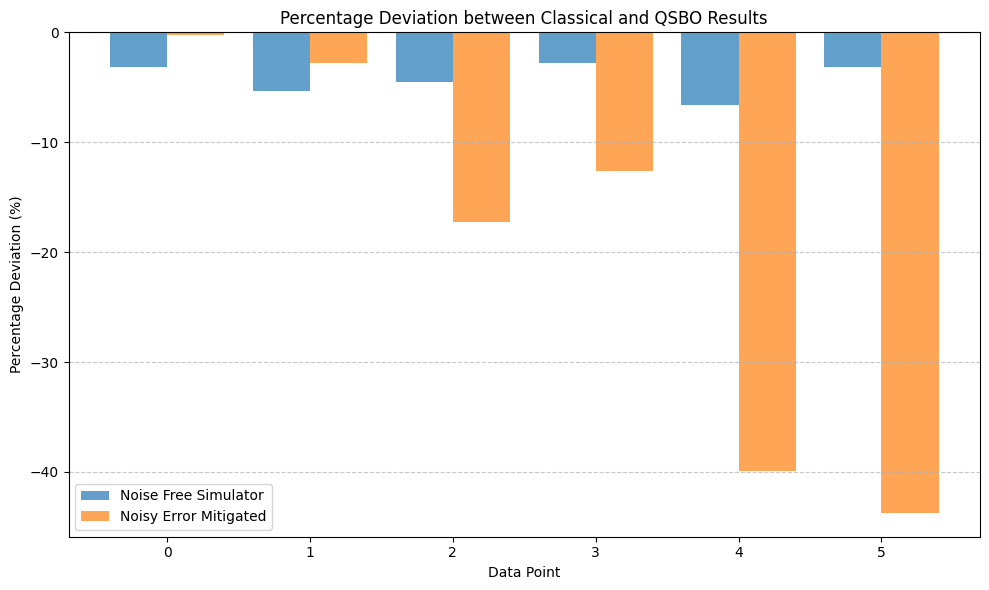}
    \caption{This plot illustrates the percentage deviation of the Maximum Expected Payoff result in both the error-free and error-mitigated backend scenarios. The baseline for comparison is the classical result. The $X-$axis represents different test instances ran for different parameters and $Y-$axis shows the percentage deviation from the ground truth result. }
    \label{fig:14}
\end{figure}

In this subsection, we experiment and report results on a second backend, a simulated environment with noise, which we seek to mitigate by applying multiple error mitigation techniques. In Fig. \ref{fig:14}, on the $x$-axis, the value $0$ denotes the original problem size, serving as a baseline. Values greater than $0$ indicate progressively increased problem sizes, with each unit increment on the x-axis corresponding to a qubit increase, which leads to increase in the demand and supply values. This labeling scheme allows for a clear comparison of the original problem size against various scaled-up versions. Despite these error mitigation strategies, the results obtained from this backend show deviations from the anticipated outcomes. This divergence (as shown in Fig.~\ref{fig:14}) highlights the influence of quantum noise and errors in practical quantum computing scenarios.

\begin{figure}[h!]
    \centering
    \includegraphics[width=\linewidth]{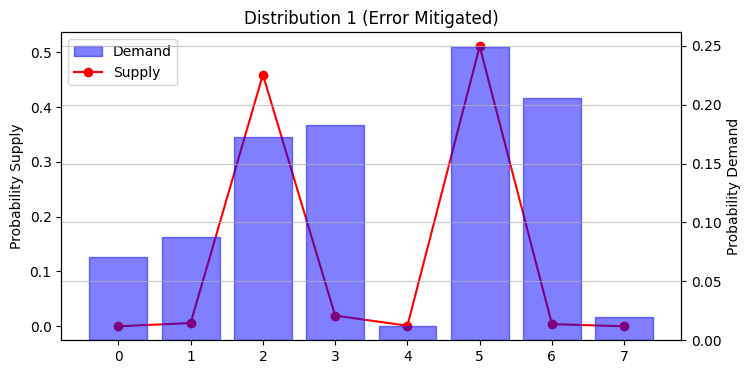}
    \caption{Optimal amount to order, run on the noisy backend with error mitigation techniques for fixed cost parameters.}
    \label{fig:9}
\end{figure}

\begin{figure}[h!]
    \centering
    \includegraphics[width=\linewidth]{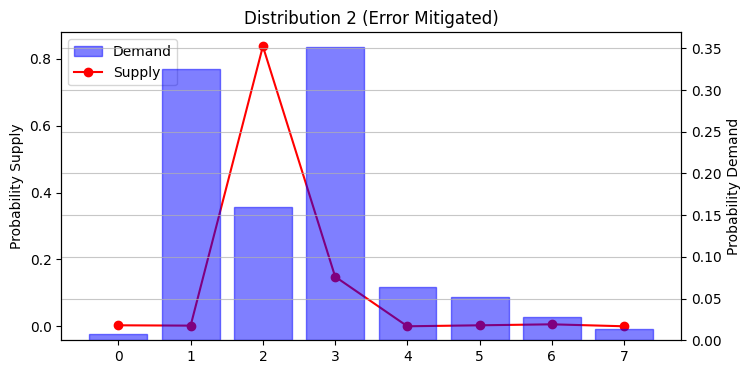}
    \caption{Optimal amount to order, run on the noisy backend with error mitigation.}
    \label{fig:11}
\end{figure}

Fig. (\ref{fig:9}) and Fig. (\ref{fig:11}) illustrate the outcomes under identical test conditions as those depicted in Fig. (\ref{fig:8}) and Fig. (\ref{fig:10}), respectively. Here, they were executed on a backend with added noise. \textbf{FakeSingapore} was used as the noisy backend, which has T1 (thermal relaxation time) and T2 (dephasing time) of $ 80.76 \mu s$ and $103.18 \mu s$ respectively.

Fig. (\ref{fig:9}) displays the optimal supply decision, revealing a dilemma between ordering $5$ and $2$, with a slightly higher likelihood assigned to ordering $5$.

In Fig. (\ref{fig:11}), the peak profit occurs at $2$, diverging from error-free outcomes (Fig. (\ref{fig:10})). Even though we applied error-mitigation techniques in the noisy backend, the effect of noise and errors is visible. While the results may closely approximate those from an error-free channel, some discrepancies are expected due to the inherent noise in the quantum system.

\subsection{Maximum Payoff Comparison}
In the following, we present a summary of results that compare:

\begin{itemize}
    \item Monte Carlo (MC) simulation, that makes use of the qGAN trained and generated data with the quantum hardware $|g_{\theta}\rangle$ and 1024 random samples of demand are drawn by measuring the distribution and used to estimate the expected payoff. This method serves as the baseline. 

    \item Our approach, i.e. QSBO, which is implemented using a quantum simulator and employing COBYLA as an optimizer with $1000$ optimization runs on $256$ quantum samples. The data utilized was generated by a qGAN and learned as $|g_{\theta}\rangle$. Within this evaluation, the Iterative Quantum Amplitude Estimation method \cite{iter} was employed, iteratively determining the optimal powers of the Grover operator and assessing the model accordingly.

    \item QSBO with similar conditions, but this time incorporating Maximum Likelihood Quantum Amplitude Estimation \cite{raymond}.
\end{itemize}

The results
for computing the maximum expected payoff are given in Table. \ref{table:1}.

\begin{table}[h!]
\centering
\caption{This table presents a comparison of different approaches to compute the expected payoff}
\begin{tabular}{ |p{2.3cm}|p{1.3cm}|p{1.5cm}|>{\raggedleft\arraybackslash}p{1.5cm}|}
 \hline
 \multicolumn{4}{|c|}{Performance Comparison} \\
 \hline
 Approach & Payoff & 95\% CI  & \#Samples\\
 \hline
 Monte Carlo & $1.00591$ & $\pm0.006925$ & $1024$ \\
 \hline
 QSBO(IQAE)  & $0.97058$ & $\pm0.008510$ &   $256$ \\
 \hline
 QSBO(MLQAE) & $0.99069$ &  $\pm0.005172$  & $256$ \\
 \hline

\end{tabular}

\label{table:1}
\end{table}

We show the 95\% confidence level (CI) for both the MC simulation and the QAE variants. Despite comparable sizes, QAE requires only a quarter of the number of samples due to better scaling. 

Note that the estimates and the CIs of the MC and the QAE evaluations are not subject to the same level comparison, while the former operates on a classical device utilizing classical samples, the latter runs on a quantum simulator, employing quantum samples defined as $2^{n}$, where $n$ represents the number of evaluation qubits or the power of the Grover operator. To run QAE on an actual quantum computer, additional improvements are required, such as long coherence times and improved gate fidelity. These are crucial for ensuring the precision and dependability of quantum computations in practical real-world scenarios.

\subsection{Scalability}

In the above instances, demand was encoded across $n=3$ qubits using qGAN modeling, accommodating values within the range of $D = [0,7]$. To encode any given maximum demand $N$, $n=\log N$ qubits are required to represent the distribution. Similarly, the decision variable $s$ is encoded within $k=\log S$ qubits ($S$ is the maximum supply amount), using the Real Amplitudes ansatz in qiskit, with a repetition count of $R=2$. Increasing the repetition layers increase the number of parameters, thereby expanding the available state space. However, this expansion simultaneously introduces greater complexity into the optimization process.\cite{express}.

The qGAN model is adaptable since it can be fitted into the complexity of the underlying data, and the resolution of the loading scheme can be traded against the complexity of the training data by adjusting the number of qubits employed and the circuit depth.  Furthermore, qGANs provide incremental learning, which means that the model can be modified as fresh training data samples become available. This can result in a significant reduction in the time during training in real-world learning scenarios.

Our approach is scalable to larger problem instances. Table \ref{Table 2} provides the scaling relationships between the quantum circuit's size and depth concerning the constituent elements of this algorithm. 
It shows that our approach scales linearly with the number of qubits, which is logarithmic in the scale of the problem. 

\begin{table}[h]  
    \centering  
    \caption{Scalability}
    \label{Table 2}
    \begin{tabular}{|c|c|c|}  
        \hline  
        Quantum Circuit & Circuit Size & Circuit Depth \\
        \hline  
        Supply & $ (2kR -k)$ & $(kR+(R-k))$ \\
        \hline
        Demand & $(2nR - n)$& $(nR+(R-n))$ \\
        \hline
        Comparison & $(10n - 9)$ & $ (13n-11)$ \\
        \hline
    \end{tabular}
\end{table}

\section{\label{sec:level8}Future Outlook}

Newsvendor Problems, previously centered on the optimization of anticipated cost and profit functions, are now venturing into more intricate terrain. This emerging frontier is characterized by the pursuit of objectives with heightened complexity, including the minimization of variance, conditional value at risk (cVaR), and other higher-order statistical metrics. These objectives, while inherently practical, present substantial computational complexities for classical computing \cite{chen2009}. 

An intriguing direction in the evolution of Newsvendor and supply chain research lies in extending traditional models to address more complex scenarios, such as Risk-Aware Procurement Optimization \cite{riskaware}. In these scenarios, multiple suppliers carry inherent risk factors that could impact procurement decisions, necessitating a more comprehensive approach to risk management within supply chain optimization frameworks. Resolving these extended Newsvendor Problems represents an engaging challenge necessitating the development of inventive algorithmic methodologies and heuristics \cite{ejor}.

In conclusion, the trajectory of Newsvendor Problems and supply chain investigation is positioned to address progressively intricate and multifaceted challenges. Those engaged in this field will be required to devise inventive optimization methodologies, confront NP-hard complexities, and incorporate different frameworks to encapsulate the intricate dynamics inherent in supply chain decision-making. As these frontiers are explored and understood, the discipline will persist in contributing insights and solutions for optimizing supply chain operations in the evolving and competitive global landscape.

\section{Acknowledgement}

This research is supported by the National Research Foundation, Singapore under its Quantum Engineering Programme 2.0 (NRF2021-QEP2-02-P01). We wish to express our heartfelt gratitude to Stefan Woerner (IBM Research Zurich) for his invaluable guidance, and Harumichi Nishimura of Nagoya University for insightful discussion and pointing out~\cite{compare}.

\appendix

\newpage

\section{Newsvendor Formulation}\label{form}

The standard newsvendor profit function is

\begin{equation}\label{eq:sum}
    \mathbb{E}[\text{profit}] = \mathbb{E}[s \text{ min}(q,X)] - cq
\end{equation}

where $X$ is a random variable with probability distribution $F$ representing demand, each unit is sold for price $s$, bought at price $c$ and $q$ is the number of units stocked. $\mathbb{E}[.]$ is the expectation operator. This model assumes fixed price and uncertain demand with limited availability and is solved using the Critical fractile formula

$$ q^{*} = F^{-1}\bigg(\frac{s-c}{s}\bigg)$$
 where $F^{-1}$ denotes the generalized inverse cumulative distribution function for $X$.

\textbf{Derivation:}

Expected value for the net revenue:

\begin{equation}
\begin{split}
    \psi(q) =& \mathbb{E}[s \text{ min}(q,X)] - cq \\
    =& \int_{0}^{\infty} [s \text{ min}(q,X) - cq] dF(x)
\end{split}
\end{equation}

Expanding on that

\begin{equation}\label{eq:A3}
    \implies \int_{0}^{\infty}  s \text{ min}(q,X) dF(x) - \int_{0}^{\infty} cq \cdot dF(x)
\end{equation}

Here the second term is just the area under the Probability Distribution Curve, i.e 1

\begin{equation*}
\begin{split}
    \int_{0}^{\infty} cq \cdot dF(x) =& cq \int_{0}^{\infty} f(x) dx \\
    =& cq \cdot 1
\end{split}
\end{equation*}

Continuing from Eq. (\ref{eq:A3}):

\begin{equation}
\begin{split}
    \implies \int_{0}^{q} s \text{ min}(q,x) dF(x) + \int_{q}^{\infty} s \text{ min}(q,x) dF(x) - cq
\end{split}
\end{equation}

In the first limit of $[0,q]$, the minimum of $(x,q) = x$, and in the second $\text{ min}(q,x) = q$ So we can rewrite as:

\begin{equation}
\begin{split}
    \implies & \int_{0}^{q} (s\cdot x) f(x) dx + \int_{q}^{\infty} (s\cdot q) dF(x) - cq \\
    = & s \int_{0}^{q} x f(x) dx + s q \int_{q}^{\infty}  dF(x) - cq \\
    = & s \int_{0}^{q} x f(x) dx + sq [F(\infty) - F(q)] - cq \\
    = & s \int_{0}^{q} x f(x) dx + sq [1 - F(q)] - cq
\end{split}
\end{equation}

Doing integration by parts for the first term:

\begin{equation}
\begin{split}
  \psi(q)  = & s[xF(x)]_{0}^{q} - s \int_{0}^{q}  f(x) dx + sq [1 - F(q)] - cq \\
    \implies & sq F(q) - s \int_{0}^{q} F(x) dx + sq (1 - F(q)) - cq \\
    \implies & sq F(q) - s \int_{0}^{q} F(x) dx + sq - sq F(q)) - cq \\
  \psi(q)  = & - s \int_{0}^{q} F(x) dx + sq -cq
\end{split}
\end{equation}

Now, looking for the maximum of the function, the maxima occur wherever the derivative of $\psi(q)$ is 0. So,

\begin{equation}
    \psi^{\prime}(q) = -sF(q) + s -c = 0
\end{equation}

This gives us:

\begin{equation}
\begin{split}
    F(q^{*}) =& 1 - \frac{c}{s} \\
    q^{*} =& F^{-1} \bigg( \frac{s-c}{s}\bigg)    
\end{split}
\end{equation}
This is our critical fractile formula, where $q^{*}$ denotes the optimum value. Also since $\psi^{''}(q) = -s f(q) < 0$ for any $q$ as the PDF $f(q)$ is non-negative, $q^{*}$ is at the maxima. 

\section{Quantum Generative Adversarial Networks}\label{quantumgan}
Loading classical data efficiently into quantum states using the best known classical methods can be done using $\mathcal{O}(2^n)$ gates into an $n$-qubit state. This scaling easily diminishes the potential quantum advantage. We use a hybrid quantum-classical algorithm for efficient, approximate quantum state loading. This loading requires $\mathcal{O}(\text{poly}(n))$ gates and can thus enable the use of potentially advantageous quantum algorithms, such as Quantum Amplitude Estimation.

The classical training data $X = \{x^0, ...,x^{s-1}\} \subset \mathbb{R}^{k_{\text{out}}}$ sampled from an unknown probability distribution $p_{\text{real}}$. Let $G_{\theta}: \mathbb{R}^{k_{\text{in}}} \rightarrow \mathbb{R}^{k_{\text{out}}}$ and $D_{\phi} : \mathbb{R}^{k_{\text{out}}} \rightarrow \{0,1\}$ denote the generator and the discriminator networks respectively. The generator translates samples from the prior distribution $p_{\text{prior}}$ into samples that are indistinguishable from the samples belonging to the real distribution $p_{\text{real}}$. The discriminator on the other hand tries to distinguish between the data from the generator and from the training set.

The generator's loss function 

\begin{equation*}
    L_G (\phi, \theta) = -\mathbb{E}_{p_{\text{prior}}}[\log (D_{\phi}(G_{\theta}(z)))]
\end{equation*}

aims at maximizing the likelihood that the generator creates samples that are labeled as real data samples. On the other hand, the discriminator's loss function

\begin{equation*}
    L_D (\phi, \theta) =  \mathbb{E}_{p_{\text{real}}} [\log D_{\phi} (x) ] +  \mathbb{E}_{p_{\text{prior}}} [1-\log (D_{\phi}(G_{\theta}(z)))]
\end{equation*}

aims at maximizing the likelihood that the discriminator labels the training data samples as training data samples and generated data samples and generated data samples. In practice, the expected values are approximated by batches of size $m$

\begin{equation*}
    {L}_{G}\left({\boldsymbol{\phi }},{\boldsymbol{\theta }}\right)=-\frac{1}{m}\sum _{l=1}^{m}\left[\log\left({D}_{{\boldsymbol{\phi }}}\left({G}_{{\boldsymbol{\theta }}}\left({z}^{l}\right)\right)\right)\right]{\rm{,and}} \\
\end{equation*}

\begin{equation*}
\begin{split}
    L_D (D_{\phi}, G_{\theta}) = \frac{1}{m} \sum_{l=1}^{m}[\log D_{\phi} (x^l) + 
    & \log(1-D_{\phi}(G_{\theta}(z^l)))]
\end{split}
\end{equation*}

for $x^l \in X$ and $z^l \approx p_{\text{prior}}$. Training the GAN is equivalent to searching a Nash equilibrium of a two-player game.

The quantum version of it uses a quantum generator and a classical discriminator to capture the probability distribution of classical training samples. In this setting, a parameterized quantum channel, i.e., the quantum generator is trained to perform the given $n-$qubit input state $|0\rangle_n$ to $n-$qubit output state:

\begin{equation}
    \mathcal{P}_X |0\rangle_n = |\psi\rangle_n = \sum_{i = 0}^{N-1} \sqrt{p_i} |i\rangle_n
\end{equation}

where $p_i$ describes the resulting occurrence probabilities of the basis states $|i\rangle_n$.

\begin{figure}
    \centering
    \includegraphics[width=\linewidth]{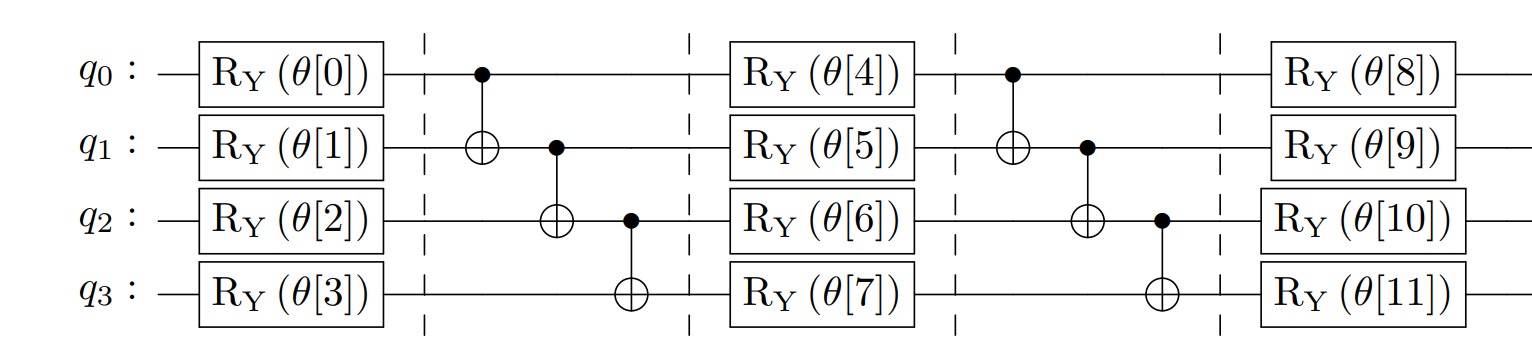}
    \caption{The variational form of the quantum generator with a depth of two and acting on four qubits. Hence it is composed of three layers of single-qubit Pauli-Y rotations and two entangling blocks in between.}
    \label{fig:enter-label}
\end{figure}

The quantum generator is implemented by a \textit{variational form} (parameterized quantum circuit). The circuit consists of a first layer of $RY$ gates and then $k$ alternating repetitions of entangling layers and further layers of $RY$ gates. The rotation acting is controlled by the parameter $\theta$, and the parameter $k$ is called the depth of the variational circuit. If such a variational circuit acts on $n$ qubits, it uses in total $(k+1)n$ parameterized single qubit gates and $kn$ two-qubit gates. The reason for choosing the $RY$ and $CZ$ gates in contrast to other Pauli rotations and two-qubit gates, is that for $\theta=0$ the variational form has no effect on the state amplitude, but only flips the phases.

The discriminator, a classical neural network, can be implemented easily using any classical machine learning library. The model is trained using COBYLA with an initial learning rate of $10^{-3}$. In each training epoch, the training data is shuffled and split into batches of size 200. the generated data samples are created by preparing and measuring the quantum generator 200 times. Then the batches are used to update the parameters of the discriminator and the generator in an alternating fashion. After the updates are completed for all batches, a new epoch starts.

\section{Value Comparison of Qubit Registers}\label{comparisonop}
The comparison scheme used in \cite{qsbo} reformulates the statement $a \ge b$ for two $n-$bit integers $a$ and $b$ to $a + (2^n-b) \ge 2^n$. 
Computing the $|2^n -b\rangle$ state first requires flipping all the $b$ states using a NOT gate and then adding the value $1$ using an adder circuit. For this, we need extra $n+1$ qubits, all initialized in state $|1\rangle$. We make efficient quantum circuits for arithmetic operations, which don't use the extra $n+1$ qubits.

\begin{figure}
    \centering
    \includegraphics[width=\linewidth]{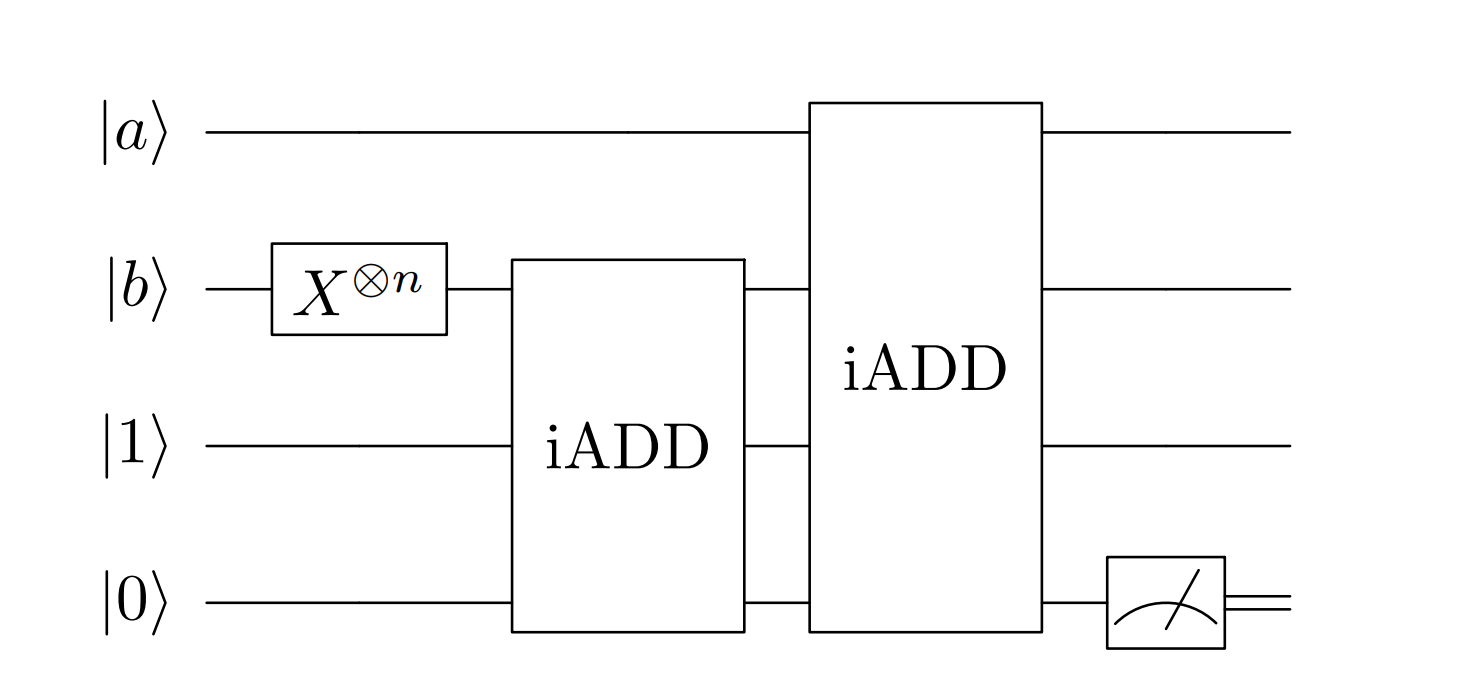}
    \caption{The circuit for comparison in \cite{qsbo}, for two bits $a$ and $b$ we need to initialize $n+1$ qubits in state $|1\rangle$ for the comparison procedure.}
    \label{fig:enter-label}
\end{figure}

Let $a$ and $b$ be two binary numbers and their respective representations are $a_{n-1} a_{n-2},..., a_{0}$ and $b_{n-1} b_{n-2},..., b_{0}$. Let $z$ be the additional qubit used to store the result. We make use of the ripple carry approach. This comparison operator is just a modification of the subtraction operator from \cite{compare}, which in turn is a slight modification of the addition operator. It effectively takes less qubit than the one mentioned in \cite{qsbo}. 

\begin{figure}[h!]
    \centering
    \includegraphics[width=\linewidth]{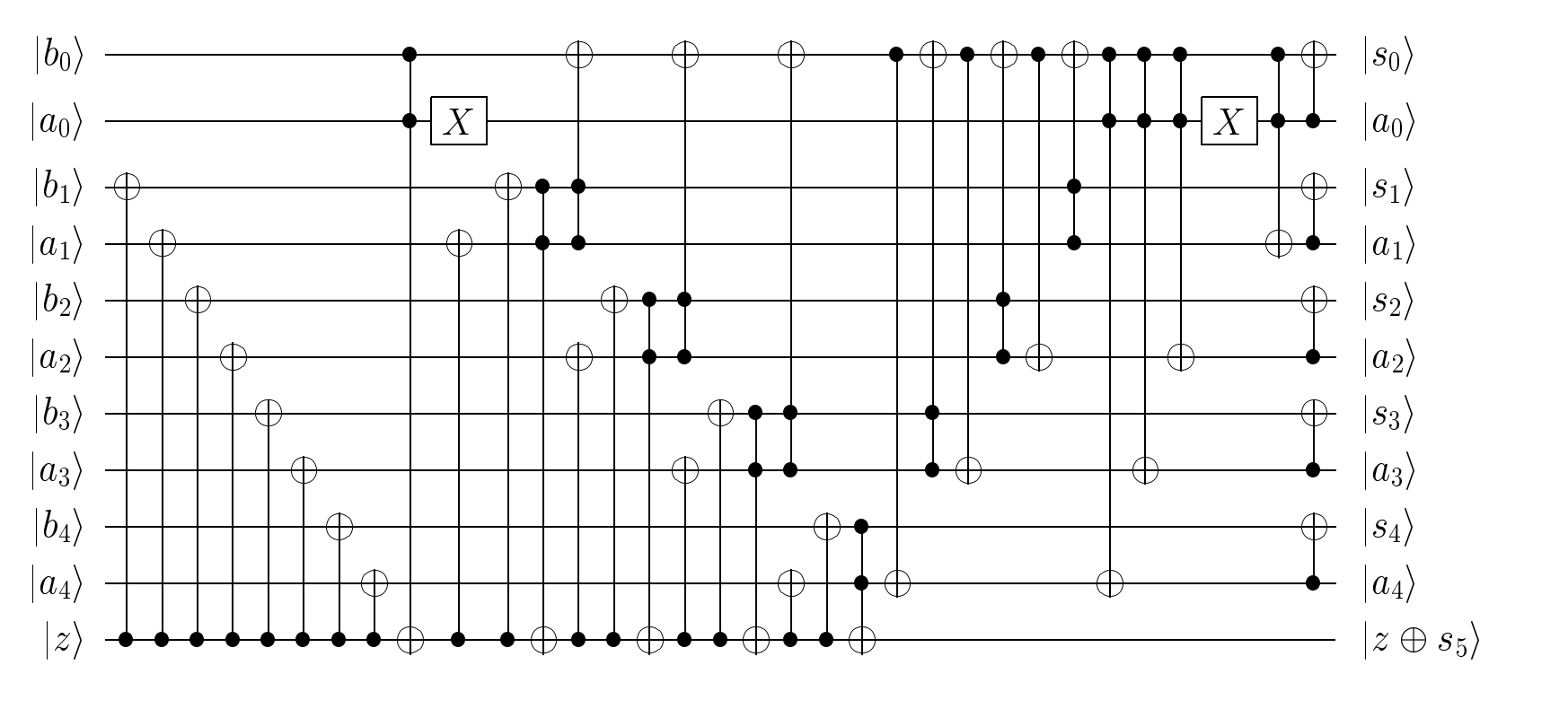}
    \caption{The circuit for comparison for $5$ digit binary.}
    \label{fig:comparison}
\end{figure}

This circuit outputs $a$, $b$ and $z \oplus y$ where $y=1$ if $b \le a $ and $0$ otherwise. The depth of this quantum circuit is $10n-9$ and the size is $13n-11$ where $n$ is the length of the binary. Although this circuit can be optimized further, the Toffoli gate can be constructed using five controlled Rotation Y gates or by using six CNOTs and eight single qubit gates. We do have other methods that can make the addition in less depth, like using the gates that are congruent to Toffoli modulo phase shifts \cite{qcqi}. It will be interesting to look at how much the circuit can be further compressed if these gates are used in place of Toffoli.

\section{{Payoff Function}}\label{payoff}

We create the operator $F$ that maps $\sum_i \sqrt{p_i} |i\rangle_n |0\rangle$ to

\begin{equation*}
    \sum_{i=0}^{2^n-1} \sqrt{p_i}|i\rangle_n \bigg[ \cos \Big(c \tilde{f}(i) + \frac{\pi}{4}\Big)|0\rangle + \sin\Big( c \tilde{f}(i) + \frac{\pi}{4}\Big) |1\rangle \bigg]
\end{equation*}

where $\tilde{f}(i)$ is just the scaled version of $f(i)$ given by:

\begin{equation}{\label{eq:d1}}
    \tilde{f}(i) = 2 \frac{f(i) - f_{min}}{f_{max} - f_{min}}-1
\end{equation}

with $f_{min} = \text{min} f(i)$ and $f_{max} = \text{max}f(i)$ and $c \in [0,1]$ is an additional scaling parameter. With these definitions, the probability of finding the ancilla qubit in state $|1\rangle$ namely

\begin{equation}
    P = \sum_{i=0}^{2^n -1} p_i \sin^2 \bigg( c \tilde{f}(i) + \frac{\pi}{4} \bigg)
\end{equation}

is well approximated by

\begin{equation}
    P \approx \sum_{i=0}^{2^n-1} p_i \bigg( c \tilde{f}(i)+ \frac{1}{2} \bigg) = c \frac{2 \mathbb{E}[f(X)] - f_{min}}{f_{max}- f_{min}} - c + \frac{1}{2}
\end{equation}

To obtain this result we made use of the approximation

\begin{equation}
    \sin^2 \bigg( c \tilde{f}(i) + \frac{\pi}{4} \bigg) = c\tilde{f}(i) + \frac{1}{2} + \mathcal{O}(c^3 \tilde{f}^3 (i))
\end{equation}

which is valid for small values of $c \tilde{f}(i)$. With this first-order approximation, the convergence rate of QAE is $\mathcal{O}(M^{-2/3})$ when $c$ is properly chosen which is already faster than the classical Monte Carlo methods \cite{risk}. We can recover the $\mathcal{O}(M^{-1})$ convergence rate of QAE by using higher orders implemented with quantum arithmetic. The resulting circuits, however, will have more gates. This trade-off is discussed in \cite{risk}, and also gives a formula that specifies which value of $c$ to use to minimize the estimation error made when using QAE.

\end{document}